\documentclass[12pt,a4paper]{article}
\makeatletter
\def\eqnarray{%
\stepcounter{equation}%
\let\@currentlabel=\theequation
\global\@eqnswtrue
\global\@eqcnt\z@
\tabskip\@centering
\let\\=\@eqncr
$$\halign to \displaywidth\bgroup\@eqnsel\hskip\@centering
$\displaystyle\tabskip\z@{##}$&\global\@eqcnt\@ne
\hfil$\displaystyle{{}##{}}$\hfil
&\global\@eqcnt\tw@$\displaystyle\tabskip\z@{##}$\hfil
\tabskip\@centering&\llap{##}\tabskip\z@\cr}
\makeatother
\newcommand{\ket}[1]{{\vert{#1}\rangle}}
\newcommand{\bra}[1]{{\langle{#1}\vert}}

\newcommand{\calh}{{\cal H}}

\newcommand{\integer}{{\mathbf Z}}
\newcommand{\fukuso}{{\mathbf C}}
\newcommand{\real}{{\mathbf R}}
\newcommand{\futon}{{\bf N}}

\newcommand{\zettai}[1]{{\vert{#1}\vert}}

\begin{document}

\title{\sl Exchange Gate on the Qudit Space and Fock Space}
\author{
  Kazuyuki FUJII
  \thanks{E-mail address : fujii@yokohama-cu.ac.jp }\\
  Department of Mathematical Sciences\\
  Yokohama City University\\
  Yokohama, 236-0027\\
  Japan
  }
\date{}
\maketitle
%\thispagestyle{empty}
%
%
%  gaiyou
%
%
\begin{abstract}
  We construct the exchange gate with small elementary gates 
  on the space of qudits, which consist of 
  three controlled shift gates and three "reverse" gates. 
  This is a natural extension of the qubit case. 
  
  We also consider a similar subject on the Fock space, but in this case 
  we meet with some different situation. However we can 
  construct the exchange gate by making use of generalized coherent 
  operator based on the Lie algebra su(2) which is a well--known 
  method in Quantum Optics. We moreover make a brief comment on 
  "imperfect clone". 
\end{abstract}
%

%\newpage

%
%
%     Honbun
%
%

\section{Introduction}

The purpose of this paper is to construct the exchange gate with 
small elementary gates on the space of qudits. 
This is a natural extension in the qubit case.

The standard model of Quantum Computation is based on $2$--level 
quantum systems (qubits) which are natural counterpart to the classical 
systems. 
In it many quantum logic gates such as controlled NOT ones, 
exchange one have been constructed, see \cite{9th}, \cite{3th} or 
\cite{KF1}, \cite{KF6}. 
The universality of quantum computation has been proved. 

\par \noindent
We have a beautiful theory, \cite{LPS}.

A major obstacle to standard quantum computation is the problem of 
decoherence depending deeply on increasing of coupled qubits. This problem 
is inevitable to every quantum system and is not easy to overcome, \cite{ASt}. 

To reduce the number of coupled qubits the non--standard model of quantum 
computation based on $d$--level quantum systems (qudits) has been considered, 
see for example \cite{CBKG}, \cite{BGS} and \cite{KBB}. This theory might  
become useful in the near future. 

However it seems (at least) to the author that there are some defects 
in this theory. Let us explain. The controlled shift gates corresponding to 
controlled NOT ones have been defined, but general controlled unitary gates 
have not been constructed. Moreover the universality on 2--qudit space 
has not been proved as far as the author knows. 

The exchange gate plays an important role in Quantum Computation. In this 
note we focus our attention to this topics. 
It is in the qubit case constructed by making use of three controlled NOT 
gates. This is a well--known result. 
We would like to consider a similar problem in the qudit case. 
In the case we can construct the exchange gate by making use of three 
controlled shift gates and three "reverse" gates. This is our main result. 

We would like to extend our problem to the case of Fock space. Namely, we 
want to construct the exchange operator with some elementary operators 
(corresponding to elementary gates in the finite--dimensional case). 
However we meet with some trouble in this line. 
We cannot define controlled shift operators as unitary ones due to the 
infinite--dimensional property. 
What elementary gates (operators) is in the infinite--dimensional case is not 
clear. 

Instead of taking all quantum states into consideration let us restrict our 
target to small set of important quantum ones, for example, 
coherent states or squeezed--coherent states which are famous in 
Quantum Optics, \cite{MW}, \cite{WPS} or \cite{KF7}. We would like to 
make good use of several methods in Quantum Optics. 
We can construct the operator exchanging coherent states by making use of 
generalized coherent operator based on Lie algebra $su(2)$ and some phase 
operators. On the other hand any quantum state can be expanded by use of 
coherent states, so we can construct the exchange operator in a perfect 
manner. 

In this section we also make a brief comment on so--called "imperfect clone" 
which is closely related to the construction of exchange gate in the case of 
coherent states, and give its general formula. 

In last let us state our hiding purpose. We are developing a geometric method 
in quantum computation called Holonomic Quantum Computation, \cite{KF00}, 
\cite{KF01}, \cite{KF02} and \cite{ZR}, \cite{PZR}. This is of course 
based on the qubit space, so we would like to generalize it on the 
qudit space. In the forthcoming papers we will report this task.

\section{Exchange Gate}

In this section we make a brief review of the construction of exchange gate 
on the qubit space 
and next extend the method from there to on the qudit space.

\subsection{Qubit Case$\cdots$ Review}

The 1--qubit space is identified with $\fukuso^{2}$ with basis
$\{\ket{0}, \ket{1} \}$ ;
\[
    \fukuso^{2} = \mbox{Vect}_{\fukuso}\{\ket{0}, \ket{1} \},\quad
    \ket{0}= {1\choose 0} ,\quad \ket{1}= {0\choose 1}\ .
\]
We consider the abelian group $\integer_{2}$ with group operation $\oplus$
\begin{equation}
\label{eq:oplus-2}
a\oplus b=a+b\ (\mbox{mod}\ 2).
\end{equation}
Explicitly
\[
   0\oplus 0=0,\ 0\oplus 1=1,\ 1\oplus 0=1,\ 1\oplus 1=0.
\]
From this it is easy to see 
\[
a\oplus b=a+b-2ab=a+(-1)^{a}b.
\]
The $t$--qubit space is the {\bf tensor product} (not direct sum) of 
$\fukuso^{2}$ 
\begin{equation}
   \label{eq:tensor-space}
   \fukuso^{2}\otimes \fukuso^{2} \otimes \cdots \otimes \fukuso^{2}
   \equiv (\fukuso^{2})^{\otimes t}
\end{equation}
with basis
\[
    \left\{
    \ket{{n_{1},n_{2}, \dots, n_{t}}} = \ket{n_{1}}\otimes \ket{n_{2}}\otimes
     \cdots \otimes \ket{n_{t}}\
     |\  n_{j} \in \integer_{2} = \{0,1 \} 
    \right\}.
\]
For example,
\[
    \ket{0}\otimes \ket{0}=
   \left(
     \begin{array}{c}
       1 \\
       0 \\
       0 \\
       0
     \end{array}
   \right),\
    \ket{0}\otimes \ket{1}=
   \left(
     \begin{array}{c}
       0 \\
       1 \\
       0 \\
       0
     \end{array}
   \right),\
    \ket{1}\otimes \ket{0}=
   \left(
     \begin{array}{c}
       0 \\
       0 \\
       1 \\
       0
     \end{array}
   \right),\
    \ket{1}\otimes \ket{1}=
   \left(
     \begin{array}{c}
       0 \\
       0 \\
       0 \\
       1
     \end{array}
   \right).
\]

\par \vspace{3mm}

Now we consider a controlled NOT operation (gate) which we will denote by
$C_{X}$ in the following. It is defined by
\begin{equation}
C_{X} : \ket{a}\otimes \ket{b}\longrightarrow 
        \ket{a}\otimes X^{a}\ket{b}=\ket{a}\otimes \ket{a\oplus b}, 
        \quad a,\ b \in \integer_{2}.
\end{equation}
Graphically it is expressed as
\begin{center}
\setlength{\unitlength}{1mm}
\begin{picture}(160,35)
\put(50,30){\line(1,0){50}}   %=89=A1=92=BC=90=FC1
\put(50,5){\line(1,0){22}}   %=89=A1=92=BC=90=FC2-1
\put(78,5){\line(1,0){22}}   %=89=A1=92=BC=90=FC2-2
\put(40,25){\makebox(9,10)[r]{$|a\rangle$}} %=8D=B6=91=A4=95=B6=8E=9A1
\put(40,0){\makebox(9,10)[r]{$|b\rangle$}} %=8D=B6=91=A4=95=B6=8E=9A2
\put(101,25){\makebox(9,10)[l]{$|a\rangle$}} %=89E=91=A4=95=B6=8E=9A1
\put(101,0){\makebox(9,10)[l]{$X^{a}\ket{b}=|a\oplus b\rangle$}}
%=89E=91=A4=95=B6=8E=9A2
\put(75,8){\line(0,1){22}}     %=8Fc=92=BC=90=FC
\put(72,25){\makebox(6,10){$\bullet$}} %=93_
\put(75,5){\circle{6}}               %=89~
\put(72,0){\makebox(6,10){X}}         %X
\end{picture}
\end{center}
and the matrix representation is
\begin{eqnarray}
   \label{eq:c-not}
C_{X} =
\left(
   \begin{array}{cccc}
        1&0&0&0 \\
        0&1&0&0 \\
        0&0&0&1 \\
        0&0&1&0
   \end{array}
\right).
\end{eqnarray}
Here we note the relation
\begin{equation}
X^{a}\ket{b}\equiv \sigma_{1}^{a}\ket{b}=\ket{a\oplus b}
\quad \mbox{for}\quad a,\ b \in \integer_{2}.
\end{equation}
where $\sigma_{1}$ is the Pauli matrx. 

\par \noindent
In this case the first bit is called a control bit and the
second a target bit. 
Of course we can consider the reverse case. Namely,
the first bit is a target one and the second a control one, which is also
called the controlled NOT operation :
\begin{equation}
\widetilde{C}_{X} : \ket{a}\otimes \ket{b}\longrightarrow 
  X^{b}\ket{a}\otimes \ket{b}=
  \ket{a\oplus b}\otimes \ket{b},\quad a,\ b \in \integer_{2},
\end{equation}
and 
graphically it is expressed as
\begin{center}
\setlength{\unitlength}{1mm}
\begin{picture}(160,35)
\put(50,5){\line(1,0){50}}   %=89=A1=92=BC=90=FC1
\put(50,30){\line(1,0){22}}   %=89=A1=92=BC=90=FC2-1
\put(78,30){\line(1,0){22}}   %=89=A1=92=BC=90=FC2-2
\put(40,25){\makebox(9,10)[r]{$|a\rangle$}} %=8D=B6=91=A4=95=B6=8E=9A1
\put(40,0){\makebox(9,10)[r]{$|b\rangle$}} %=8D=B6=91=A4=95=B6=8E=9A2
\put(101,25){\makebox(9,10)[l]{$X^{b}\ket{a}=|a\oplus b\rangle$}} %
\put(101,0){\makebox(9,10)[l]{$|b\rangle$}}
%=89E=91=A4=95=B6=8E=9A2
\put(75,5){\line(0,1){22}}     %=8Fc=92=BC=90=FC
\put(72,0){\makebox(6,10){$\bullet$}} %=93_
\put(75,30){\circle{6}}               %=89~
\put(72,25){\makebox(6,10){X}}        %X
\end{picture}
\end{center}
and the matrix representation is
\begin{eqnarray}
   \label{eq:c-hat-not}
\widetilde{C}_{X} =
\left(
   \begin{array}{cccc}
        1&0&0&0 \\
        0&0&0&1 \\
        0&0&1&0 \\
        0&1&0&0
   \end{array}
\right).
\end{eqnarray}

\par \vspace{5mm} \noindent
The exchange (swap) gate is defined by 
\begin{equation}
\label{eq:exchange-gate}
 S(\ket{a}\otimes \ket{b})=\ket{b}\otimes \ket{a}
 \quad \mbox{for}\quad a,\ b \in \integer_{2}.
\end{equation}
It is well--known that 
this gate is constructed by making use of three controlled 
NOT gates 
\begin{equation}
\label{eq:exchange-formula}
 S=C_{X}\circ \widetilde{C}_{X}\circ C_{X}
\end{equation}
and is graphically expressed as
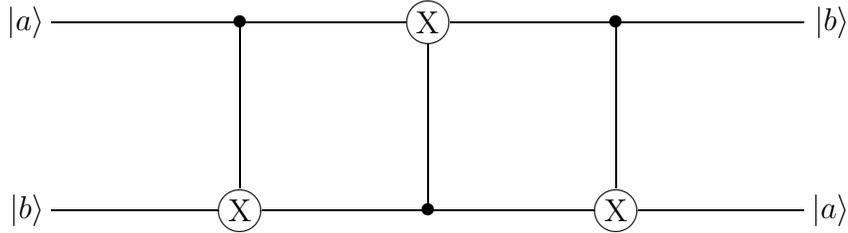
\begin{center}
\setlength{\unitlength}{1mm}
\begin{picture}(200,35)
\put(30,30){\line(1,0){47}}   %=89=A1=92=BC=90=FC1
\put(30,5){\line(1,0){22}}   %=89=A1=92=BC=90=FC2-1
\put(58,5){\line(1,0){22}}   %=89=A1=92=BC=90=FC2-2
\put(55,8){\line(0,1){22}}     %=8Fc=92=BC=90=FC
\put(20,25){\makebox(9,10)[r]{$|a\rangle$}} %=8D=B6=91=A4=95=B6=8E=9A1
\put(20,0){\makebox(9,10)[r]{$|b\rangle$}} %=8D=B6=91=A4=95=B6=8E=9A2
\put(52,25){\makebox(6,10){$\bullet$}} %=93_
\put(55,5){\circle{6}}               %=89~
\put(52,0){\makebox(6,10){X}}         %X
\put(80,5){\line(1,0){22}}   %=89=A1=92=BC=90=FC1
\put(83,30){\line(1,0){22}}   %=89=A1=92=BC=90=FC2-1
\put(108,30){\line(1,0){22}}   %=89=A1=92=BC=90=FC2-2
\put(108,5){\line(1,0){22}}   %=89=A1=92=BC=90=FC2-2
\put(80,5){\line(0,1){22}}     %=8Fc=92=BC=90=FC
\put(77,0){\makebox(6,10){$\bullet$}} %=93_
\put(80,30){\circle{6}}               %=89~
\put(77,25){\makebox(6,10){X}}        %X
\put(105,30){\line(1,0){22}}   %=89=A1=92=BC=90=FC1
\put(108,5){\line(1,0){22}}   %=89=A1=92=BC=90=FC2-2
\put(105,8){\line(0,1){22}}     %=8Fc=92=BC=90=FC
\put(127,25){\makebox(9,10)[r]{$|b\rangle$}} %=8D=B6=91=A4=95=B6=8E=9A1
\put(127,0){\makebox(9,10)[r]{$|a\rangle$}} %=8D=B6=91=A4=95=B6=8E=9A2
\put(102,25){\makebox(6,10){$\bullet$}} %=93_
\put(105,5){\circle{6}}               %=89~
\put(102,0){\makebox(6,10){X}}         %X
\end{picture}
\end{center}

\subsection{Qudit Case}

The 1--qudit space is identified with $\fukuso^{d}$ with basis
$\{\ket{0}, \ket{1}, \cdots, \ket{d-1}\}$ ;
\[
  \fukuso^{d} = \mbox{Vect}_{\fukuso}\{\ket{0}, \ket{1}, \cdots, \ket{d-1}\},
  \quad
  \ket{j}= (0,\cdots,0,1,0,\cdots,0)^{T}\ (j-\mbox{position}), 
\]
where $T$ is the transpose of a vector. 

\par \noindent 
We consider the abelian group $\integer_{d}$ with group operation $\oplus$
\begin{equation}
\label{eq:oplus-3}
a\oplus b=a+b\ (\mbox{mod}\ d).
\end{equation}

\par \noindent 
Here we deal with two elementary operations $\{\Sigma_{1}, \Sigma_{3}\}$ in 
the unitary group 
$U(d)$ which correspond to the Pauli matrices $\{\sigma_{1}, \sigma_{3}\}$. 
They are defined as 
\begin{equation}
\Sigma_{1}\ket{a}=\ket{a\oplus 1},\quad 
\Sigma_{3}\ket{a}=\zeta^{a}\ket{a} \quad \mbox{for}\quad a \in \integer_{d}, 
\end{equation}
where $\zeta=\mbox{exp}(2\pi{i}/d)$. We note that they are not hermitian 
\[
\Sigma_{1}^{\dagger}=\Sigma_{1}^{d-1}, \quad 
\Sigma_{3}^{\dagger}=\Sigma_{3}^{d-1}, \quad \mbox{and}\quad 
\Sigma_{3}\Sigma_{1}=\zeta\Sigma_{1}\Sigma_{3}.
\]
In the following we set $\Sigma=\Sigma_{1}$ for simplicity. 

\par \vspace{3mm} \noindent
The controlled shift gates are defined as 
\begin{equation}
\label{eq:controlled-shift-1}
 C_{\Sigma}(\ket{a}\otimes \ket{b})=\ket{a}\otimes {\Sigma}^{a}\ket{b}
   =\ket{a}\otimes \ket{a\oplus b}
\end{equation}
and 
\begin{equation}
\label{eq:controlled-shift-2}
 \widetilde{C}_{\Sigma}(\ket{a}\otimes \ket{b})
   ={\Sigma}^{b}\ket{a}\otimes \ket{b}
   =\ket{a\oplus b}\otimes \ket{b}. 
\end{equation}

\par \noindent 
We  note that 
$
C_{\Sigma}(\ket{a}\otimes \ket{0})=\ket{a}\otimes \ket{a}
$ 
if we set $b=0$ in (\ref{eq:controlled-shift-1}). That is, the basis 
$\{\ket{a}\ |\ 0\leq a \leq d-1\}$ can be cloned.

\par \vspace{5mm} \noindent
The exchange (swap) gate which is our target in this paper is given by 
\begin{equation}
\label{eq:general-definition}
 S=\sum_{a,b=0}^{d-1}(\ket{a}\otimes \ket{b})(\bra{b}\otimes \bra{a}) 
  =\sum_{a,b=0}^{d-1}\ket{a}\bra{b}\otimes \ket{b}\bra{a}.
\end{equation}

\par \noindent
{\bf Problem}\quad How can we construct this gate with some elementary 
gates like (\ref{eq:exchange-formula}) ?

\par \vspace{5mm} \noindent
Before dealing with this problem 
let us define an important operation $K$ in $U(d)$ :
\begin{equation}
\label{eq:K-gate}
 K(\ket{a})=\ket{a\oplus a\oplus \cdots \oplus a \ (d-2\ \mbox{times})}
 \quad \mbox{for}\quad 
 a \in \integer_{d}.
\end{equation}
Since 
\[
  a\oplus a\oplus \cdots \oplus a
  =\left\{
   \begin{array}{ll}
     \displaystyle{0}\ \qquad \  \mbox{if}\quad a=0  \\
     \displaystyle{d-a}\quad  \mbox{if}\quad a\neq 0
   \end{array}
 \right.
\],
we have $K(\ket{0})=\ket{0}$ and $K(\ket{a})=\ket{d-a}$ for $1\leq a \leq 
d-1$. This is a kind of reverse operation. 
We note that when $d=2$ this operation is just the identity, so 
we cannot see it in the figure. 

\par \vspace{3mm} \noindent 
Now we are in a position to state our main result.\ 
The exchange (swap) gate is constructed as 
\begin{equation}
\label{eq:general-exchange-gate}
 S=C_{\Sigma}\circ (K\otimes 1)\circ \widetilde{C}_{\Sigma}
   \circ (K\otimes 1)\circ C_{\Sigma}\circ (1\otimes K)
\end{equation}
and it is graphically expressed as
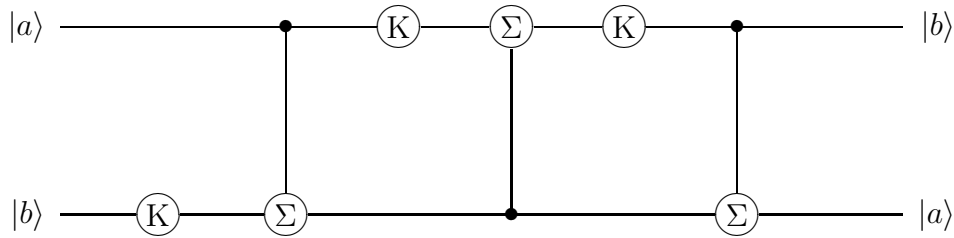
\begin{center}
\setlength{\unitlength}{1mm}
\begin{picture}(200,35)
\put(25,30){\line(1,0){42}}   %=89=A1=92=BC=90=FC1
\put(25,5){\line(1,0){10}}   %=89=A1=92=BC=90=FC2-1
\put(41,5){\line(1,0){11}}   %=89=A1=92=BC=90=FC2-1
\put(58,5){\line(1,0){27}}   %=89=A1=92=BC=90=FC2-2
\put(55,8){\line(0,1){22}}     %=8Fc=92=BC=90=FC
\put(38,5){\circle{6}}               
\put(35,0){\makebox(6,10){K}}         
\put(14,25){\makebox(9,10)[r]{$|a\rangle$}} %=8D=B6=91=A4=95=B6=8E=9A1
\put(14,0){\makebox(9,10)[r]{$|b\rangle$}} %=8D=B6=91=A4=95=B6=8E=9A2
\put(52,25){\makebox(6,10){$\bullet$}} %=93_
\put(55,5){\circle{6}}               %=89~
\put(52,0){\makebox(6,10){$\Sigma$}}         %X
\put(70,30){\circle{6}}               
\put(67,25){\makebox(6,10){K}}  
\put(73,30){\line(1,0){9}}   %=89=A1=92=BC=90=FC1      
\put(85,5){\line(1,0){27}}   %=89=A1=92=BC=90=FC1
\put(88,30){\line(1,0){9}}   %=89=A1=92=BC=90=FC2-1
\put(103,30){\line(1,0){12}}   %=89=A1=92=BC=90=FC2-2
%\put(108,5){\line(1,0){22}}   %=89=A1=92=BC=90=FC2-2
\put(85,5){\line(0,1){22}}     %=8Fc=92=BC=90=FC
\put(82,0){\makebox(6,10){$\bullet$}} %=93_
\put(85,30){\circle{6}}               %=89~
\put(82,25){\makebox(6,10){$\Sigma$}}        
\put(100,30){\circle{6}}               
\put(97,25){\makebox(6,10){K}}  
\put(115,30){\line(1,0){22}}   %=89=A1=92=BC=90=FC1
\put(118,5){\line(1,0){19}}   %=89=A1=92=BC=90=FC2-2
\put(115,8){\line(0,1){22}}     %=8Fc=92=BC=90=FC
\put(135,25){\makebox(9,10)[r]{$|b\rangle$}} %=8D=B6=91=A4=95=B6=8E=9A1
\put(135,0){\makebox(9,10)[r]{$|a\rangle$}} %=8D=B6=91=A4=95=B6=8E=9A2
\put(112,25){\makebox(6,10){$\bullet$}} %=93_
\put(115,5){\circle{6}}               %=89~
\put(112,0){\makebox(6,10){$\Sigma$}}         %X
\end{picture}
\end{center}

\par \noindent
Our result is rather complicated compared with (\ref{eq:exchange-formula}).

\subsection{Problem}

Here let us present an important problem related to the universality of
2--qudit spaces. 

\par \noindent
We usually assume that in the quantum computation
\begin{itemize}
 \item[(1)] each qudit space is universal, 
 \item[(2)] controlled shift gates can be constructed (efficiently).
\end{itemize}
Here the terminology "universal" means that any unitary element in 
$U(d)$ can be constructed. 

\par \noindent
Now we define controlled unitary gates 
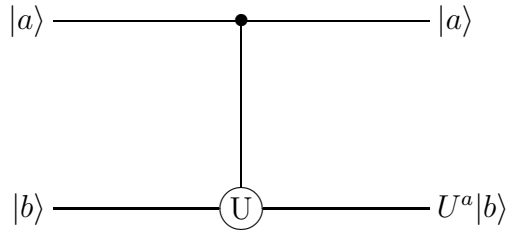
\begin{equation}
\label{eq:controlled-unitary}
 C_{U}(\ket{a}\otimes \ket{b})=\ket{a}\otimes U^{a}\ket{b}, 
\end{equation}
where $U \in U(d)$. 

\begin{center}
\setlength{\unitlength}{1mm}
\begin{picture}(160,35)
\put(50,30){\line(1,0){50}}   %=89=A1=92=BC=90=FC1
\put(50,5){\line(1,0){22}}   %=89=A1=92=BC=90=FC2-1
\put(78,5){\line(1,0){22}}   %=89=A1=92=BC=90=FC2-2
\put(40,25){\makebox(9,10)[r]{$|a\rangle$}} %=8D=B6=91=A4=95=B6=8E=9A1
\put(40,0){\makebox(9,10)[r]{$|b\rangle$}} %=8D=B6=91=A4=95=B6=8E=9A2
\put(101,25){\makebox(9,10)[l]{$\ket{a}$}} %=89E=91=A4=95=B6=8E=9A1
\put(101,0){\makebox(9,10)[l]{$U^{a}\ket{b}$}}
%=89E=91=A4=95=B6=8E=9A2
\put(75,8){\line(0,1){22}}     %=8Fc=92=BC=90=FC
\put(72,25){\makebox(6,10){$\bullet$}} %=93_
\put(75,5){\circle{6}}               %=89~
\put(72,0){\makebox(6,10){U}}         %X
\end{picture}
\end{center}

Another controlled unitary gates $\widetilde{C}_{U}$ defined by 
$
\widetilde{C}_{U}(\ket{a}\otimes \ket{b})=U^{b}\ket{a}\otimes \ket{b}
$ 
are obtained easily 
\begin{equation}
S\circ C_{U}\circ S=\widetilde{C}_{U}
\end{equation}
by making use of the exchange gate constructed above. 

\par \vspace{3mm} \noindent
{\bf Problem (A)}\ Under the above assumption (1) and (2) can 
all controlled unitary gates (operations) be constructed ? 

\par \vspace{3mm} \noindent
This is a crucial problem. Moreover 

\par \vspace{3mm} \noindent
{\bf Problem (B)}\ Is the 2--qudit space universal ? or 
Can every unitary operation in $U(d^2)$ be constructed under the 
assumption ? 

\par \vspace{3mm} \noindent
As far as we know these problems have not been solved, \cite{KFu}.

\section{Exchange Operator in the Fock Space}

We make a short review of general theory of both a coherent operator and 
generalized coherent one based on Lie algebra $su(2)$, and explain our 
result and present the related problems. 

Let $a(a^\dagger)$ be the annihilation (creation) operator of the harmonic 
oscillator.
If we set $N\equiv a^\dagger a$ (:\ number operator), then
\begin{equation}
  \label{eq:3-1}
  [N,a^\dagger]=a^\dagger\ ,\
  [N,a]=-a\ ,\
  [a^\dagger, a]=-\mathbf{1}\ .
\end{equation}
Let $\calh$ be a Fock space generated by $a$ and $a^\dagger$, and
$\{\ket{n}\vert\  n\in\futon\cup\{0\}\}$ be its basis.
The actions of $a$ and $a^\dagger$ on $\calh$ are given by
\begin{equation}
  \label{eq:3-2}
  a\ket{n} = \sqrt{n}\ket{n-1}\ ,\
  a^{\dagger}\ket{n} = \sqrt{n+1}\ket{n+1}\ ,
  N\ket{n} = n\ket{n}
\end{equation}
where $\ket{0}$ is a normalized vacuum ($a\ket{0}=0\  {\rm and}\  
\langle{0}\vert{0}\rangle = 1$). From (\ref{eq:3-2})
state $\ket{n}$ for $n \geq 1$ are given by
\begin{equation}
  \label{eq:3-3}
  \ket{n} = \frac{(a^{\dagger})^{n}}{\sqrt{n!}}\ket{0}\ .
\end{equation}
These states satisfy the orthogonality and completeness conditions 
\begin{equation}
  \label{eq:3-4}
   \langle{m}\vert{n}\rangle = \delta_{mn}
   \quad \mbox{and}\quad 
   \quad \sum_{n=0}^{\infty}\ket{n}\bra{n} = \mathbf{1}\ . 
\end{equation}
We call a state defined by 
\begin{equation}
\label{eq:3-5}
\ket{z} = \mbox{e}^{za^{\dagger}- \bar{z}a}\ket{0}\equiv D(z)\ket{0} 
\quad \mbox{for}\quad z\ \in\ \fukuso 
\end{equation}
the coherent state.

Next let us define a squeezed operator $S(w)$. 
If we set 
\begin{equation}
  \label{eq:3-6}
  K_{+}\equiv{1\over2}\left( a^{\dagger}\right)^2\ ,\
  K_{-}\equiv{1\over2}a^2\ ,\
  K_{3}\equiv{1\over2}\left( a^{\dagger}a+{1\over2}\right)\ ,
\end{equation}
then it is easy to check the $su(1,1)$ relations 
\begin{equation}
  \label{eq:3-7}
 [K_{3}, K_{+}]=K_{+}, \quad [K_{3}, K_{-}]=-K_{-}, 
 \quad [K_{+}, K_{-}]=-2K_{3}.
\end{equation}
That is, the set $\{K_{+},K_{-},K_{3}\}$ gives a unitary representation of 
$su(1,1)$ with spin $K = 1/4\ \mbox{and}\ 3/4$, \cite{AP}. 
We call an operator 
\begin{equation}
  \label{eq:3-8}
   S(w) = \mbox{e}^{wK_{+}-\bar{w}K_{-}}=
   \mbox{e}^{\frac{1}{2}\{w(a^{\dagger})^2 - \bar{w}a^2\}}
   \quad \mbox{for} \quad w \in \fukuso 
\end{equation}
the squeezed operator.

\par \noindent 
The operators $D(z)$ and $S(w)$ play a crucial role in Quantum Optics, 
\cite{MW} and \cite{WPS}. 

Here let us define a squeezed--coherent state 
\begin{equation}
  \label{eq:squeezed-coherent}
   \ket{(w,z)}=S(w)D(z)\ket{0}\quad \mbox{for}\quad w,\ z \in \fukuso. 
\end{equation}
This state plays very important role in Holonomic Quantum Computation 
\cite{KF00}, \cite{KF01}, \cite{KF02}.

\par \vspace{5mm} \noindent
Next we consider the system of two--harmonic oscillators. If we set
\begin{equation}
  \label{eq:3-9}
  a_1 = a \otimes 1,\  {a_1}^{\dagger} = a^{\dagger} \otimes 1;\ 
  a_2 = 1 \otimes a,\  {a_2}^{\dagger} = 1 \otimes a^{\dagger},
\end{equation}
then it is easy to see 
\begin{equation}
  \label{eq:3-10}
 [a_i, a_j] = [{a_i}^{\dagger}, {a_j}^{\dagger}] = 0,\ 
 [a_i, {a_j}^{\dagger}] = \delta_{ij}\quad \mbox{for} \quad 
 i, j = 1, 2. 
\end{equation}
We also denote by $N_{j} = {a_j}^{\dagger}a_j$ ($j=1, 2$) number operators.

Now we can construct representation of Lie algebras $su(2)$ by 
making use of Schwinger's boson method, see \cite{FKSF1}, \cite{FKSF2}. 
Namely if we set 
\begin{equation}
  \label{eq:3-11}
     J_+ = {a_1}^{\dagger}a_2,\ J_- = {a_2}^{\dagger}a_1,\ 
     J_3 = {1\over2}\left({a_1}^{\dagger}a_1 - {a_2}^{\dagger}a_2\right), 
\end{equation}
then it is easy to check that (\ref{eq:3-11}) satisfies the $su(2)$ 
relations 
\begin{equation}
  \label{eq:3-12}
 [J_{3}, J_{+}]=J_{+}, \quad [J_{3}, J_{-}]=-J_{-}, 
 \quad [J_{+}, J_{-}]=2J_{3}.
\end{equation}
In the following we define (unitary) generalized coherent operators 
based on Lie algebras $su(2)$. 
We set 
\begin{equation}
  \label{eq:3-13}
  U_{J}(z) = e^{z{a_1}^{\dagger}a_2 - \bar{z}{a_2}^{\dagger}a_1}\quad 
  {\rm for}\quad  z \in \fukuso .
\end{equation}
This operator has an important property : $U_{J}(z)\ket{0}\otimes \ket{0}
=\ket{0}\otimes \ket{0}$. That is, this keeps the vacuum invariant. 
For the details of $U_{J}(z)$ see \cite{AP} or \cite{FKSF1}, \cite{FKSF2}. 

\par \vspace{3mm} \noindent
Here we define unitary phase operators 
\begin{equation}
  \label{eq:3-14}
V_{j}(\theta) = e^{i\theta N_{j}}\quad {\rm for}\quad \theta \in \real ,
\end{equation}
then we can add a phase to the parameter of coherent states like 
$
\ket{\alpha}_{j}\longrightarrow \ket{\mbox{e}^{i\theta}\alpha}_{j}
$ 
by making use of (\ref{eq:3-14}). For the details see \cite{KF7}. 

\par \vspace{3mm} 
Let us make some mathematical preliminaries for the latter. 
We have easily 
\begin{eqnarray}
  \label{eq:J-rotation-1}
  U_{J}(t)a_{1}U_{J}(t)^{-1}&=&
      cos(\zettai{t})a_{1}-\frac{tsin(\zettai{t})}{\zettai{t}}a_{2}, \\
  \label{eq:J-rotation-2}
  U_{J}(t)a_{2}U_{J}(t)^{-1}&=&
      cos(\zettai{t})a_{2}+\frac{{\bar t}sin(\zettai{t})}{\zettai{t}}a_{1},
\end{eqnarray}
so the map 
$
(a_{1},a_{2}) \longrightarrow 
          (U_{J}(t)a_{1}U_{J}(t)^{-1},U_{J}(t)a_{2}U_{J}(t)^{-1}) 
$
is 
\[
(U_{J}(t)a_{1}U_{J}(t)^{-1},U_{J}(t)a_{2}U_{J}(t)^{-1})=
(a_{1},a_{2})
\left(
  \begin{array}{cc}
     cos(\zettai{t})& \frac{{\bar t}sin(\zettai{t})}{\zettai{t}}\\
     -\frac{tsin(\zettai{t})}{\zettai{t}}& cos(\zettai{t})
   \end{array}
 \right)(\in SU(2)).
\]

\par \vspace{3mm} \noindent
The exchange operator $S : \calh\otimes \calh \longrightarrow 
\calh\otimes \calh$ is given by 
\begin{equation}
\label{eq:infinite-definition}
 S=\sum_{a,b=0}^{\infty}(\ket{a}\otimes \ket{b})(\bra{b}\otimes \bra{a}) 
  =\sum_{a,b=0}^{\infty}\ket{a}\bra{b}\otimes \ket{b}\bra{a}.
\end{equation}
However this is nothing but a mathematical formula. 

\par \vspace{3mm} \noindent
{\bf Problem}\quad How can we construct this operator with some elementary 
ones ?

\par \vspace{3mm} \noindent
Here we meet with some trouble. Let us explain. 
Even in this case we can define a shift operator 
\[
\Sigma : \calh \longrightarrow \calh, \quad 
\Sigma(\ket{a})=\ket{a+1}\quad \mbox{for}\quad a\geq 0. 
\]
But this operator is not unitary ! \ Namely, we cannot realize a controlled 
shift operator like $C_{\Sigma}$ within unitary operators. 

\par \vspace{3mm} \noindent
{\bf Problem}\quad What is an operator corresponding to the controlled 
shift gate in the finite--dimensional cases ? 

\par \vspace{3mm} \noindent
From now on let us change our purpose. We consider not all quantum states 
but small and important quantum states like coherent ones or 
squeezed--coherent ones.

The coherent states in (\ref{eq:3-5}) are exchanged by "elementary gates" 
in Quantum Optics 
\begin{equation}
  \label{eq:coherent-exchange}
     \ket{z_{1}}\otimes \ket{z_{2}} \longrightarrow 
     \ket{z_{2}}\otimes \ket{z_{1}}.
\end{equation}
The outline of the proof is as follows \cite{KF7}. 
By making use of $U_{J}(t)$ and (\ref{eq:J-rotation-1}), 
(\ref{eq:J-rotation-2}) 
we can show 
\begin{equation}
\label{eq:general-formula}
\ket{z_{1}}\otimes \ket{z_{2}}\  \longrightarrow \ 
\ket{
cos(|t|)z_{1}+\mbox{e}^{i\theta}sin(|t|)z_{2}
}
\otimes
\ket{
cos(|t|)z_{2}-\mbox{e}^{-i\theta}sin(|t|)z_{1}
} 
\end{equation}
where $t=|t|\mbox{e}^{i\theta}$. This is a crucial formula. 
By setting for example $|t|=\pi/2$ 
\begin{equation}
\label{eq:general-formula-restriction}
\ket{z_{1}}\otimes \ket{z_{2}}\  \longrightarrow \ 
\ket{\mbox{e}^{i\theta}z_{2}}\otimes \ket{-\mbox{e}^{-i\theta}z_{1}} 
=
\ket{\mbox{e}^{i\theta}z_{2}}\otimes 
\ket{\mbox{e}^{-i(\theta+\pi)}z_{1}} 
\end{equation}
and removing the phases by making use of $V_{j}$ ($j=1, 2$) at (\ref{eq:3-14}) 
we finally obtain (\ref{eq:coherent-exchange}). 
Let us note that our operators $U_{J}(t)$ and $V_{j}(\theta)$ don't depend on 
both $z_{1}$ and $z_{2}$. 

\par \vspace{3mm} \noindent 
Moreover we can perform what we have called "imperfect clone" for the 
coherent states 
\begin{equation}
  \label{eq:coherent-clone}
     \ket{z}\otimes \ket{0} \longrightarrow 
     \ket{\mbox{cos}(|t|)z}\otimes \ket{\mbox{sin}(|t|)z} 
\end{equation}
by (\ref{eq:general-formula}). These are a satisfactory result. 

Now we are in a position to construct the general exchange gate. 
Let $\ket{X}$ and $\ket{Y}$ be any quantum states in the Fock space $\calh$ 
and we show 
\begin{equation}
  \label{eq:general formula of exchage}
  \ket{X}\otimes \ket{Y}\longrightarrow  \ket{Y}\otimes \ket{X}.
\end{equation}
Let us recall that the typical feature of coherent states $\ket{z}\ 
(z \in \fukuso)$ in (\ref{eq:3-5}) is the resolution of unity 
\[
   \int_{\fukuso}\frac{[d^{2}z]}{\pi}\ket{z}\bra{z}
  =\sum_{n=0}^{\infty}\ket{n}\bra{n}={\bf 1}.
\]
See for example \cite{KF7}. From this 
\begin{equation}
  \label{eq:expansion-formula}
  \ket{M}=\int_{\fukuso}\frac{[d^{2}z]}{\pi}{\langle{z}|{M}\rangle}\ket{z}
  \quad \mbox{for}\quad M=X,\ Y, 
\end{equation}
so we have 
\begin{equation}
\ket{X}\otimes \ket{Y}=
\int\int\frac{[d^{2}z]}{\pi}\frac{[d^{2}w]}{\pi}
{\langle{z}|{X}\rangle}{\langle{w}|{Y}\rangle}\ket{z}\otimes \ket{w}. 
\end{equation}
However we have shown $\ket{z}\otimes \ket{w}\longrightarrow 
\ket{w}\otimes \ket{z}$ by (\ref{eq:coherent-exchange}), so that 
\begin{equation}
\int\int\frac{[d^{2}w]}{\pi}\frac{[d^{2}z]}{\pi}
{\langle{w}|{Y}\rangle}{\langle{z}|{X}\rangle}\ket{w}\otimes \ket{z}
=
\ket{Y}\otimes \ket{X}. 
\end{equation}
We finally obtain (\ref{eq:general formula of exchage}). This is our 
main result in this section.

\par \vspace{5mm} \noindent

In last we state the general theory of our "imperfect clone". 
Let $\ket{X}$ be any element in the Fock space $\calh$. Then 
by (\ref{eq:expansion-formula}) we can write as 
\[
\ket{X}\otimes \ket{0}=\int\frac{[d^{2}z]}{\pi}{\langle{z}|{X}\rangle}
\ket{z}\otimes \ket{0}.
\]
On the other hand we have constructed the "imperfect clone" 
$
\ket{z}\otimes \ket{0}\longrightarrow 
\ket{\mbox{cos}(|t|)z}\otimes \ket{\mbox{sin}(|t|)z}
$
by (\ref{eq:coherent-clone}), so 
\begin{equation}
  \label{eq:target-formula}
\ket{X}\otimes \ket{0}\longrightarrow 
\int\frac{[d^{2}z]}{\pi}{\langle{z}|{X}\rangle}
\ket{\mbox{cos}(|t|)z}\otimes \ket{\mbox{sin}(|t|)z}.
\end{equation}
We have only to calculate the integral in the right hand side. 
The coherent state $\ket{z}$ in (\ref{eq:3-5}) is expanded as 
\begin{equation}
\ket{z}=\mbox{e}^{-|z|^{2}/2}\sum_{n=0}^{\infty}\frac{z^n}{\sqrt{n!}}\ket{n}, 
\end{equation}
see for example \cite{KF7}. By expressing $\ket{X}$ as 
\[
\quad \ket{X}=\sum_{n=0}^{\infty}x_{n}\ket{n}\quad \mbox{for}\quad 
x_{n}\in \fukuso
\]
and after some algebras we have
\[
\mbox{RHS of (\ref{eq:target-formula})}=
\sum_{n,m=0}^{\infty}\sqrt{\frac{(n+m)!}{n!m!}}
\mbox{cos}^{n}(|t|)\mbox{sin}^{m}(|t|)x_{n+m}\ket{n}\otimes \ket{m}.
\]
That is, we finally obtain the general formula of "imperfect clone" 
\begin{equation}
  \label{eq:imperfect-clone-formula}
\ket{X}\otimes \ket{0}\longrightarrow 
\sum_{n,m=0}^{\infty}\sqrt{\frac{(n+m)!}{n!m!}}
\mbox{cos}^{n}(|t|)\mbox{sin}^{m}(|t|)x_{n+m}\ket{n}\otimes \ket{m}.
\end{equation}
In particular when $\mbox{cos}(|t|)=\mbox{sin}(|t|)=1/\sqrt{2}$ we have 
\begin{equation}
\ket{X}\otimes \ket{0}\longrightarrow 
\sum_{n,m=0}^{\infty}\sqrt{\frac{(n+m)!}{n!m!}}2^{-(n+m)/2}
x_{n+m}\ket{n}\otimes \ket{m}.
\end{equation}

A comment is in order. The author doesn't know whether this general 
formula of "imperfect clone" is really useful or not in Quantum Information 
Theory. A further study is needed.

\par \vspace{10mm} \noindent
{\it Acknowledgment.}
The author wishes to thank Yoshinori Machida for his warm hospitality 
at Numazu College of Technology and Kunio Funahashi for useful discussions.

%%%%%%%%%%%%%
%References%
%%%%%%%%%%%%%

\end{document}